\documentclass[aps,pra,reprint,floatfix,superscriptaddress,amssymb,amsmath]{revtex4-1}
\usepackage{graphicx}
\usepackage{hyperref}
\usepackage{braket}
\usepackage{color}

\newcommand{\diff}{\mathop{}\!\mathrm{d}}
\newcommand{\comma}{\text{,}}

\newcommand{\point}{\text{.}}
\newcommand{\real}{\text{Re}}
\newcommand{\imag}{\text{Im}}
\newcommand{\inb}{\text{in}}
\newcommand{\out}{\text{out}}

\begin{document}

\title{Quantum Simulation with a Boson Sampling Circuit}

\author{Diego Gonz\'{a}lez \surname{Olivares}}
\affiliation{Instituto de F\'{i}sica Fundamental IFF-CSIC, Calle Serrano 113b, Madrid E-28006, Spain.}
\author{Borja \surname{Peropadre}}
\affiliation{Department of Chemistry and Chemical Biology, Harvard University, Cambridge, Massachusetts 02138, United States}
\author{Al\'an \surname{Aspuru-Guzik}}
\affiliation{Department of Chemistry and Chemical Biology, Harvard University, Cambridge, Massachusetts 02138, United States}
\author{Juan Jos\'e \surname{Garc\'{i}a-Ripoll}}
\affiliation{Instituto de F\'{i}sica Fundamental IFF-CSIC, Calle Serrano 113b, Madrid E-28006, Spain.}

\begin{abstract}
In this work we study a system that consists of $2M$ matter qubits that interact through a boson sampling circuit, i.e., an $M$-port interferometer, embedded in two different architectures. We prove that, under the conditions required to derive a master equation, the qubits evolve according to effective bipartite XY spin Hamiltonians, with or without local and collective dissipation terms. This opens the door to the simulation of any bipartite spin or hard-core boson models and exploring dissipative phase transitions as the competition between coherent and incoherent exchange of excitations. We also show that in the purely dissipative regime this model has a large number of exact and approximate dark states, whose structure and decay rates can be estimated analytically. We finally argue that this system may be used for the adiabatic preparation of boson sampling states encoded in the matter qubits.
\end{abstract}

\maketitle

\section{\label{sec:Introduction}Introduction}

Optical circuits are linear devices that route photons along different paths. They can be found at a variety of scales, from classical circuits built using macroscopic lenses and mirrors\ \cite{Reck1994a,Clements2016a}, to photonic crystals that achieve routing and confinement by means of nanostructuring a metamaterial\ \cite{Lund-Hansen2008a,Goban2014a}, on-chip waveguides\ \cite{Politi2008a}, waveguides imprinted using femtosecond pulses\ \cite{Meany2015a} or reconfigurable optical microchips\ \cite{Carolan2015a,harris2015}.

In addition to their widespread use in telecommunications, optical circuits have found two extraordinary applications in quantum science. Photonic pathways enable the engineering of electromagnetic environments for atoms or quantum dots, either to to enhance and control light-matter interactions\ \cite{Goban2014a,Lodahl2015a}, implement nonlinear transformations on light\ \cite{Tiecke2014a}, or engineer photon-mediated interactions\ \cite{Chang2006a,Gonzalez-Tudela2011a}.

An additional novel application is the study of computational models based on boson sampling\ \cite{Aaronson2011a, gard15, broome2013, spring2013, tillmann2013, crespi2013,Shen2014,motes2014,peropadre2015a,motes2015linear,huh2015}. Within this paradigm, a particular kind of optical circuit known as a multiport interferometer is fed with a nonclassical input of one photon in $N$ out of $M\gg N$ ports. Aaronson and Arkhipov showed\ \cite{Aaronson2011a} that the events with exactly $N$ photons exiting $M$ distinct ports have a probability distribution that, under reasonable conjectures, is \emph{classically hard} to simulate for arbitrary circuits\ \footnote{When $U$ is sampled randomly from the set of unitaries using the Haar measure}. Therefore, optical circuits are good candidates to demonstrate the supremacy of quantum computing models\ \cite{Preskill2012a}.

\begin{figure}[t!] 
\centering \includegraphics[width=\linewidth]{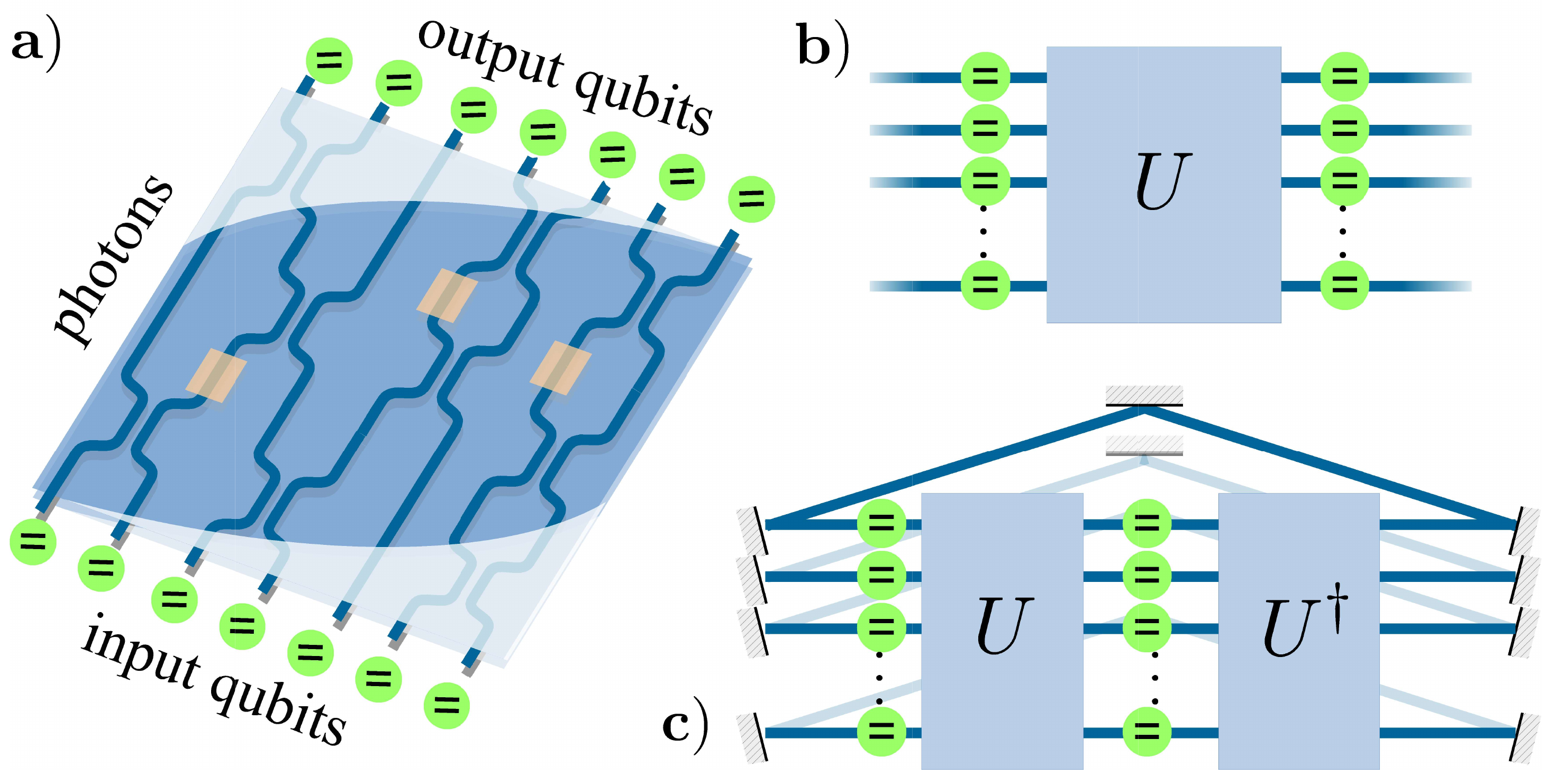}
\caption{(a) Our setup consists of two sets of matter qubits or two-level systems (green) effectively connected with each other by the photons that propagate through a linear optics circuit (blue waveguides). Such a circuit could be built. for instance, from beam splitters and phase shifters. This circuit may be open (b), with free photons, or it may be converted into a resonator by terminating the ends with mirrors or closing it periodically (c) with additional optical paths.}
\label{fig:Setup}
\end{figure}

In this work, we merge the two research lines mentioned above into an application that studies the long time effective dynamics between matter qubits interacting through \emph{boson sampling circuits} [cf.\ Fig.\ \ref{fig:Setup}a]. We build on the idea that the time evolution with general spin XY Hamiltonians is formally linked to boson sampling\ \cite{Peropadre2015b}. Here, we show that this relation emerges not only at a mathematical level, but \emph{also} in physical implementations. More precisely, we consider a setup with $2M$ two-level systems coupled to the ports of an $M$-line interferometer ---symbolized by green circles and blue waveguides in Fig.\ \ref{fig:Setup}a---. We show that this system exhibits an effective spin-spin interaction which may be dissipative [cf.\ Fig.\ \ref{fig:Setup}b] or coherent [cf.\ Fig.\ \ref{fig:Setup}c]. In the first case, the dissipative interaction has got a collective nature. We prove that these models have a large space of dark and quasi-dark states, which we can analyze for arbitrary random unitaries. In the second case, our system implements  arbitrary bipartite $XY$ spin models with long-range interactions, whose dynamics is classically hard to simulate. Moreover, the same setup can be used to directly prepare and study boson sampling states directly in the matter qubits. The main conclusion of this work is that the combination of optical circuits and few-level systems offers unique opportunities for quantum simulation and quantum information processing.

The structure of this work is as follows. Throughout Sec.\ \ref{sec:Physical-Setup} we introduce the physical setup and its description in terms of abstract unitary transformations. Starting from these \emph{spin-boson models}, Sec.\ \ref{sec:Effective-Models} derives the effective interaction between emitters, both in the coherent and incoherent regimes. These models are then used in Sec.\ \ref{sec:Applications} to study applications in quantum simulations, the potential of achieving quantum supremacy and dissipative state engineering. Sec.\ \ref{sec:Conclusions} summarizes the results and discusses their potential impact in various fields. Finally, for the sake of readability, we group several appendices with the explicit calculations for the interferometer transformation, the effective models and the boson-samplign state preparation.

\section{\label{sec:Physical-Setup}Physical Setup}

We consider two arrays of qubits or two-level systems that interact through an optical circuit. The circuit is regarded as a linear transformation $U$ of the annihilation operators, from $M$ input channels to $M$ output channels,
\begin{equation}
a^\prime_{mk} = U^{}_{mn} \! \left(k\right) a^{}_{nk} \, \comma
\end{equation}
where the linear transformation depends in general on the linear momentum, $k$, associated to the bosonic modes with anihilation operators $a^{}_{mk}$, $a^\prime_{mk}$. As explained in Appendix~\ref{app:Unitary}, this unitary map can be built using beam splitters and phase shifters. Nevertheless, the same idea can be extended to more general setups with optical\ \cite{tillmann2013,broome2013,spring2013,motes2014} or microwave media\ \cite{peropadre2015a} that propagate photons through a finite number of channels.

We consider an architecture in which we have one matter qubit coupled to each of the input or output ports of the photonic circuit [cf. Fig.~\ref{fig:Setup}(a)]. This architecture may be embedded in two different physical configurations. In the first of them, shown in Fig.~\ref{fig:Setup}(b), the photonic channels extend in both directions well beyond the qubits and support propagating waves. The Hamiltonian reads
\begin{align} \label{eq:H-Open}
H = & \sum\limits_{m=1}^M \frac{\Delta}{2} \left( \sigma^z_{\inb,m} + \sigma^z_{\out,m} \right)
+ \sum\limits_{m=1}^M \sum\limits_k \omega_k \> a^\dagger_{mk} a^{}_{mk}
\nonumber\\
+ & \sum\limits_{m,k} g_k \, \sigma^x_{\inb,m}
\left( a^\dagger_{mk} + a^{}_{mk} \right) +\\
+ & \hspace{-0.35em} \sum\limits_{m,n,k} \hspace{-0.3em} g_k \, \sigma^x_{\out,m}
\left( U^{}_{mn} \! \left(k\right) a^\dagger_{nk}
+ U^*_{mn} \! \left(k\right) a^{}_{nk} \right) \comma \nonumber
\end{align}
where $\Delta$ is the qubit splitting, $\omega_k$ is the photon frequency, $k$ the corresponding momentum and the $\sigma^{\beta}_{\alpha,m}$ are the Pauli operators for the two sets of $\alpha=$ in,out of $M$ matter qubits each. The qubit-mode coupling constant is of the form $g_k = \bar{g}_k/\sqrt{L} = \mu \sqrt{2\pi\omega_k}/\sqrt{L}$, with $L$ being a quantization length for the waveguide modes and $\mu$ the dipolar coupling strength. We will eventually take the continuum limit replacing the sum over momenta with an integral, $\sum_k g^2_k\to\frac{1}{2\pi}\int_{-\infty}^{+\infty}\diff k \, \bar{g}^2_k$, but the sums are kept for convenience throughout the calculations.

Note also the difference in the coupling amplitudes of the input and ouptut qubits in Eq.~\eqref{eq:H-Open}. The output qubits couple through the unitary transformation $U \! \left(k\right)$ implemented by the optical circuit, which in general is a function of the photon momentum $k$ with the only constraint $U \! \left(-k\right)=U^* \! \left(k\right)$ [cf. Appendix~\ref{app:Unitary}].

An alternative setup would be an optical circuit where the output ports are closed with mirrors, thereby creating a resonator. The analysis of such circuits might be complicated in general, because the unitary $U$ depends on photon momentum and the modes need to satisfy zero-field boundary conditions. For the sake of simplicity we have devised a configuration of the form shown in Fig.~\ref{fig:Setup}(c). This configuration ensures that we may define photonic modes provided that $k L^\prime = 2\pi z$, where $z\in\mathbb{Z}$ and $L^\prime$ is the total length of the resonator. In addition, input and output qubits perceive the same distribution of fields in this configuration as they do in the setup of Fig.~\ref{fig:Setup}(b). Consequently, the Hamiltonian of this alternative configuration is the same as \eqref{eq:H-Open}, with the difference that we will never replace the sums with integrals.

\section{\label{sec:Effective-Models}Effective Models}

So far we have considered linear transformations on a collection of bosonic modes connecting two sets of qubits. We will now derive effective models for the qubits by tracing out those bosonic degrees of freedom in the two different setups considered. We begin with the resonator system, in which the discrete spectrum gives rise to a purely coherent interaction. We then continue with the open waveguide circuit, for which both Hamiltonian interactions and collective dissipation coexist.

\subsection{\label{sec:Closed-Circuit}Closed Circuit: Spin Hamiltonian}

We will work with the model for the resonator setup shown in Fig.~\ref{fig:Setup}(c), where the optical circuit is introduced twice to have appropriate boundary conditions. We assume a dispersive limit in which the frequency spacing between cavity modes $\delta\omega = c \, 2\pi/L$ is much larger than the qubit-resonator coupling $g_k\sim g$. In this regime, if qubits are off-resonant from all cavity modes and $\left\vert \omega_k-\Delta \right\vert > g$, we can use second-order perturbation theory to derive an effective qubit interaction mediated by the exchange of virtual photons
\begin{align} \label{eq:H-Spin}
H_\text{spin} = & \sum\limits_{m=1}^M \left( \frac{\tilde{\Delta}}{2} \, \sigma^z_{\inb,m}
                                          + \frac{\tilde{\Delta}}{2} \, \sigma^z_{\out,m} \right)
\\ + & \sum\limits_{m,n} J^{}_{mn}
       \left( \sigma^+_{\out,m} \sigma^-_{\inb,n} + \sigma^+_{\inb,m} \sigma^-_{\out,n} \right)
\point \nonumber
\end{align}
The virtual photon exchange gives rise to a renormalization of the qubit frequency $\tilde{\Delta} = \Delta \, + \, \delta$, where
\begin{equation}
\delta = \sum\limits_k \frac{g^2_k}{\Delta - \omega_k} \, \comma
\end{equation}
is the effective shift; and the effective exchange interaction is given by
\begin{equation}\label{eq:EffectiveCouplingResonator}
J^{}_{mn} = \sum\limits_k \real \left[ U^{}_{mn} \left(k\right) \right] \frac{g^2_k}{\Delta - \omega_k}
\, \comma
\end{equation}
which depends on the transformation $U\left(k\right)$ implemented by the optical circuit for each value of the photon momentum $k$. In most cases, one of the contributions in~\eqref{eq:EffectiveCouplingResonator} will dominate with respect to all the others, allowing a direct identification of $J^{}_{mn}$ with $\real \left[ U^{}_{mn} \left(\omega_k\approx\Delta\right) \right]$.

\subsection{\label{sec:Open-Circuit}Open Circuit: Master Equation}

We will now work with the model~\eqref{eq:H-Open}, in which the photons form a continuum of modes propagating in both directions. Following the derivation in Appendix~\ref{app:Master-Eq-Derivation}, we obtain an effective master equation for the reduced density matrix of the qubits $\rho_0$. This equation only depends on the unitary transformation that represents the optical circuit at the resonance point, $U\equiv U \! \left( k_\Delta \right)$, and the spontaneous emission rate $\Gamma$ of each qubit onto its corresponding photonic channel [cf. Eq.~\ref{eq:SpontaneousEmissionRate}]:
\begin{equation}\label{eq:Effective-Master-Eq}
\frac{\diff\rho_0}{\diff t} = -i \, [H_\text{eff},\rho_0] + \Gamma \, \mathcal{L}[\rho_0].
\end{equation}
This equation contains an effective Hamiltonian
\begin{equation}\label{eq:Effective-H}
H_\text{eff} = \Gamma \sum_{m,n} \imag \left[ U_{mn} \right] \, \sigma_{\out,m}^+ \, \sigma_{\inb,n}^-
                               + \text{h.c.} \comma
\end{equation}
and a dissipation term
\begin{align} \label{eq:Dissipation-Terms}
\mathcal{L}[\rho_0] &
= \sum\limits_m \hspace{-0.25em} \sum\limits_{\alpha \in \lbrace\inb,\out\rbrace} \hspace{-1em}
                \mathcal{L}_1 \left[ \rho_0; \sigma^-_{\alpha,m}, \sigma^+_{\alpha,m} \right]
\\
& + \sum\limits_{m,n} \real \left[ U^{}_{mn} \right]
    \mathcal{L}_1 \left[ \rho_0; \sigma^+_{\out,m}, \sigma^-_{\inb,n} \right] + \text{h.c.} \comma
\nonumber
\end{align}
defined in terms of the Lindblad superoperator
\begin{equation}
\mathcal{L}_1 \left[ \rho_0; A, B \right] =
B \rho_0 A - \frac{1}{2} \left\lbrace \rho_0, AB \right\rbrace \point
\end{equation}
The dissipation terms in Eq.~\eqref{eq:Dissipation-Terms} are consistent with the application of Fermi's golden rule to the degrees of freedom of this system. The $\real \left[U^{}_{nm}\right]$ factors accompanying the nonlocal spin-flip terms arise from the fact each spin can excite photons in both directions, $k>0$ and $k<0$, and that the dependence on $k$ of the corresponding matrix elements for the associated transitions satisfies $U(-k)=U(k)^*$ [cf. Appendix~\ref{app:Unitary}].

\section{\label{sec:Applications}Applications}

\subsection{\label{sec:Bipartite-XY-Simulation}Quantum Simulation of Spin Models and Spin-Boson Sampling}

Working in the resonator regime, Eq.~\eqref{eq:H-Spin} opens the door to the simulation of \emph{any} bipartite spin or hard-core boson model. More precisely, for any bipartite spin-spin interaction described by a real and symmetric matrix $J$, we can identify a unitary matrix $U$ such that $J\propto\real\left[U\right]$ in an element-wise manner. The procedure for this would start by diagonalizing $J=W^\dagger\Lambda W$, for a certain unitary transformation $W$ and a diagonal form $\Lambda_{mn}=\lambda_m\delta_{mn}$. We then would find out the largest eigenvalue $\delta =\max |\Lambda_{mm}|$ and construct $U=W^\dagger \exp \left( i\Theta \right) W$, where the diagonal matrix $\Theta_{mn}=\theta_{m}\delta_{mn}$ is chosen such that $\cos(\theta_m) = \lambda_m/\delta$.

A very relevant subset of problems in this context corresponds to \emph{spin-sampling}. In this case $U\in SO(M)\in\mathbb{R}^{M\times M}$ would be a random orthogonal matrix drawn from the Haar measure. As it was proven in Ref.~\cite{Peropadre2015b}, an $XY$ model with random, long-range interactions implements a short-time dynamics that is as complex as boson sampling. Our resonator setup provides a possible physical implementation of this idea. More precisely, if we excite $N\ll M$ input spins and probe the output qubits after a time $T\simeq \pi/\delta$, the distribution of excitations in this subsystem would be described by a permanent, just as in the case of boson sampling. Provided that $M$ is large enough, the resulting dynamics would be classically hard to simulate.

\subsection{\label{sec:Dark-States}Dissipative Regime and Dark States}

It is also interesting to take the opposite limit in which coherent tunneling is completely suppressed and we only have collective dissipation. In this case, $U=O$, where $O$ is an orthogonal transformation, and we can write the master equation as
\begin{equation} \label{eq:Factorized-Master-Equation}
\frac{\diff\rho_0}{\diff t} = \Gamma \, \mathcal{L}_O \left[ \rho_0 \right] =
\Gamma \sum\limits_{m=1}^M \mathcal{L}_1 \left[ \rho_0; S^-_m, S^+_m \right] \comma
\end{equation}
in terms of collective spin operators
\begin{equation} S^+_m =
\frac{1}{\sqrt{2}} \left( \sigma^+_{\inb,m} + \sum\limits_n \, O^{}_{nm} \, \sigma^+_{\out,n}  \right).
\end{equation}
We may now look for stationary states, solving the equation $\mathcal{L}_O \left[ \rho_0 \right]=0$. Besides the trivial stationary solution that is the ground state $\ket{0} = \otimes^{2M}_{m=1} \ket{\downarrow_m}$, we will find $M$ exact \emph{`dark states'} $W^+\ket{0}$ of the dynamics that correspond to delocalized spin excitations. These states, which are created by the operators
\begin{equation}
\label{eq:Dark-States}
W^+_m = \frac{1}{\sqrt{2}}
\left( \sum\limits_n O^{}_{mn} \sigma^+_{\inb,n} - \sigma^+_{\out,m} \right),
\end{equation}
are called dark states because they are exactly decoupled from the photonic fields. The states \eqref{eq:Dark-States} appear as a generalization of the singlet states\ \cite{Gonzalez-Tudela2011a} that are the dark states of a system consisting of two qubits interacting with a lossless photonic waveguide.

\begin{figure}[t!]
\centering \includegraphics[width=\linewidth]{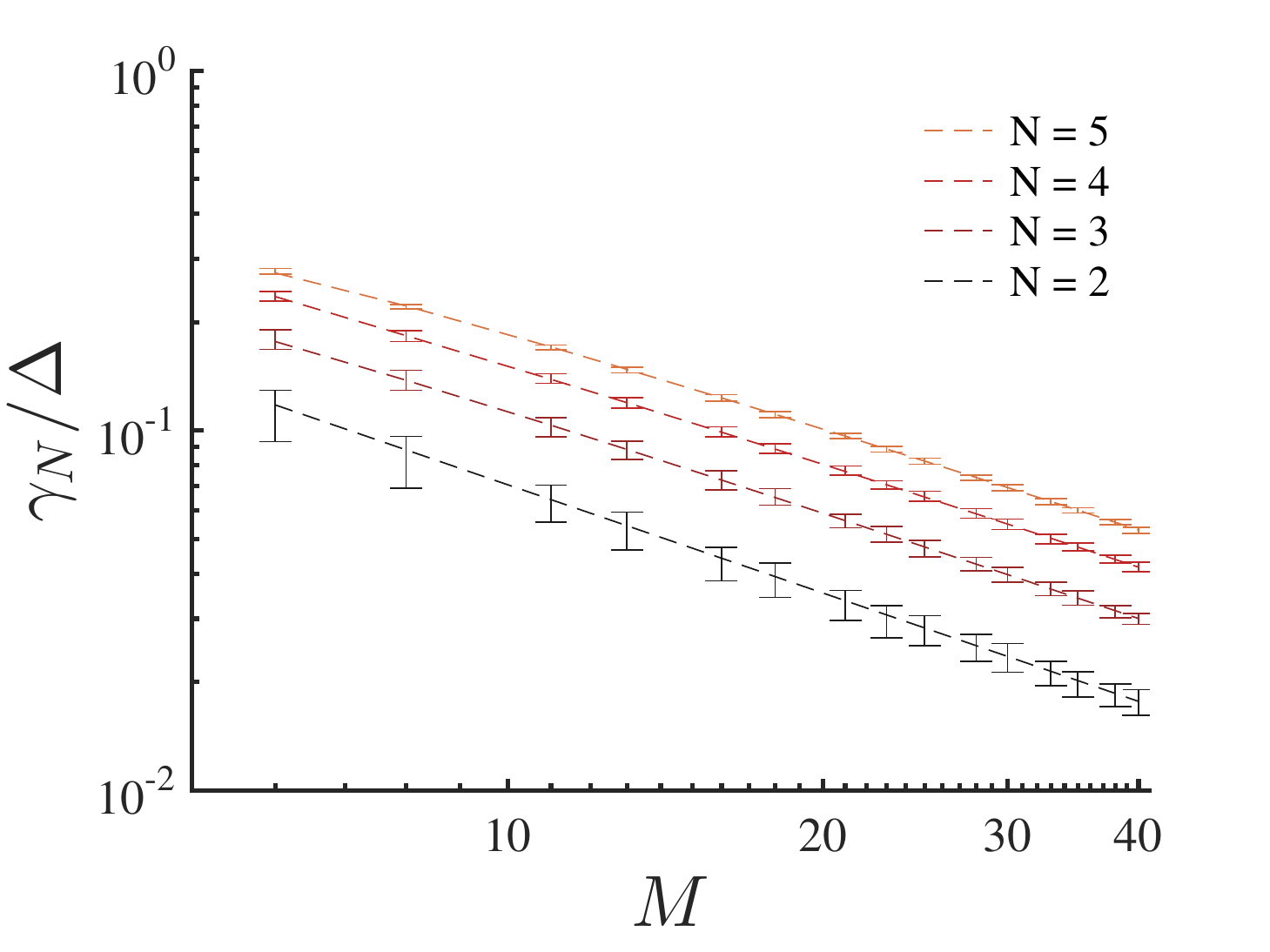}
\caption{Effective decay rates averaged over random unitary operators $U$ for different number of modes $M$
and quasiparticle filling $N=2,\ldots 5$. Dashed lines represent the analytic results~\eqref{eq:CrowdedRates} for increasing values of $N$ (from bottom to top), while the corresponding sets of points with error bars represent the decay rates estimated from a Monte Carlo simulation of the system.}
\label{fig:Decay}
\end{figure}

In addition to these \emph{exact} dark states, in our problem we also find other \emph{quasi-stationary} states that are constructed by repeatedly applying different $W^+_m$ operators. As explained in Appendix~\ref{app:Dark-State-Decays}, we find that a state with $N$ distinct dark-state quasiparticle excitations decays with a rate
\begin{equation} \label{eq:CrowdedRates}
\gamma^{}_N = \Gamma \,
\sqrt{ \sum_{n=2}^N \frac{\left(-1\right)^n}{2^{n+1}} \frac{N!}{\left(N-n\right)!}\frac{n!}{M^n} }
\, \comma
\end{equation}
with $\gamma^{}_1 :=0$. Thus, in the very dilute limit, $\gamma^{}_N$ can be very small and the resulting states may be regarded as \emph{de facto} dark states. This limit of diluteness is reached even for moderate circuit sizes. We have verified this performing a Monte Carlo simulation of the circuit and estimating the decay rates with up to $M=50$ modes and up to $N=5$ excitations. These results are shown in  Fig.~\ref{fig:Decay} alongside the theoretical predictions.

\subsection{\label{sec:Adiabatic-boson sampling}Adiabatic Preparation of Boson Sampling States}

An attractive feature of model~\eqref{eq:H-Spin} is that it can be used to adiabatically prepare the boson sampling state. Given a random unitary transformation $U$ sampled with the Haar measure, the boson sampling state with $N$ excitations in $M$ modes is given by
\begin{equation}
\ket{\phi_{\mathrm{BS}}} =
\prod_{n=1}^N \sum_{m=1}^M U^{}_{mn} \, a_m^\dagger\ket{0} \point
\end{equation}
This state is the one obtained by injecting $N$ bosons in the first $N\ll M$ modes of a multiport interferometer implementing the transformation $U$. We now show how to encode an approximately similar state in the qubits
\begin{equation}
\ket{\psi_{\mathrm{BS}}} = \prod_{n=1}^N \sum_{m=1}^M U^{}_{mn} \, \sigma_{out,m}^+\ket{0}
                         + \mathcal{O}\left(\varepsilon\right) \comma
\end{equation}
where the error $\varepsilon = \psi-\phi$ can be made arbitrarily small.

Our protocol builds on the results of~\cite{Peropadre2015b}, which states that the dynamics of a multimode bosonic system with few excitations $(N\ll M)$ can be approximated by the evolution of a spin model with a similarly small number of excitations. The protocol assumes that we can build a  random spin-spin interaction of the form\ \eqref{eq:H-Spin} where $J = U \in \mathbb{R}^{M\times M}$ is our randomly sampled orthogonal transformation. We also assume that in the effective model we can tune the energies of the input and output qubits, using external fields. The time-dependent Hamiltonian reads
\begin{equation}
H = H_\text{spin} + \epsilon \sum\limits_{m=1}^M
                             \left( \left[ 1 - \lambda\left(t\right) \right] \sigma^z_{\out,m}
                                                           + \lambda\left(t\right) \sigma^z_{\inb,m} \right)
\point
\end{equation}
The switching function $\lambda\left(t\right)$ interpolates smoothly $\lambda\left(0\right) = 1$ and $\lambda\left(T\right) = 0$ over a long time $T$. We start with an initial state $\ket{\psi\left(0\right)} = \prod_{m=1}^{N}\sigma^+_{\inb,m}\ket{0}$, in a regime in which $\epsilon \, \lambda\left(0\right) \gg \left\vert J \right\vert$ prevents tunneling. We then adiabatically shift $\lambda$ to zero, until at a time $T$ we have $\lambda\left(T\right) = 0$. As shown in Appendix~\ref{app:boson sampling}, provided $\dot{\lambda}$ remains small compared to $\left\vert J \right\vert$, the system should then converge to a ground state that consists on $N$ excitations in the output qubits. In other words, $\ket{\psi(T)}\simeq \ket{\psi_{\mathrm{BS}}}$. The resulting state should be arbitrarily close to a boson sampling state, provided that the number of excitations is dilute enough.

\section{\label{sec:Conclusions}Conclusions}

We have presented a setup and a model for engineering photon-mediated interactions between two-level emitters using optical circuits. This idea represents a rather general paradigm that encompasses and extends previous approaches towards similar goals in one-dimensional photonic environments\ \cite{Chang2006a,Gonzalez-Tudela2011a}. Using the tools in this paper we can reverse-engineer arbitrary bipartite interactions, such as high-dimensional XY spin Hamiltonians, finding the optical circuits that implement them, and rely on reconfigurable circuits\ \cite{Carolan2015a} or single-purpose devices\ \cite{Politi2008a,Meany2015a} to implement them.

We have discussed various applications of the resulting setups, that range from studies of quantum complexity and quantum supremacy at through short time evolution\ \cite{Peropadre2015b} or through the preparation of boson sampling states, to using the optical circuit dark states for quantum information and quantum optics applications.

These applications can be tested in a variety of state-of-the-art platforms. For instance, setups with trapped atoms in photonic crystals have demonstrated strong light-matter interactions\ \cite{Tiecke2014a} that are sufficient for implementing the dissipative models in this work. Solid-state devices such as quantum dots have also achieved sufficient coupling strengths\ \cite{Lodahl2015a}, but in this case inhomogeneous broadening of levels might make them more suitable for studying disorder in our spin Hamiltonians.

All our proposals can be extended to work with superconducting quantum circuits, where microwave transformations such as beam splitters have been demonstrated\ \cite{Hoffmann2010a,Pechal2016a}. In this case, the enhanced light-matter interaction allows reaching the ultrastrong coupling regime to the continuum\ \cite{Forn-Diaz2016a} and we can no longer apply the Markov approximations. However, rather simple generalizations of our treatment based on the polaron transformation\ \cite{Kurcz2014a,Diaz-Camacho2015a} shows that we still recover spin-spin interactions, but now they become of Ising type. This opens the door to simulating other types of dissipative phase transitions~\cite{Rossini07,Fazio00,Keeling12,Eisert15}, but also opens questions regarding the quantum complexity of Ising models and their time evolution.

Another important generalization would be using only a subset of qubits, or placing qubits at a subset of ports and blocking other channels with mirrors or closed loops. These and other designs, which allow implementing more general spin Hamiltonians which are not bipartite, will be explored in further work. 

Finally, this work has been developed under reasonable assumptions of Markovianity and long photon wavepackets, where the time for photons to travel between qubits greatly exceeds the spontaneous emission rate. These open interesting questions about how to generalize our theoretical framework to include retardation effects.

\acknowledgments

The authors acknowledge support from the European Union FP7 project PROMISCE, Spanish MINECO projects FIS2012-33022 and FIS2015-70856-P, CAM Research Network QUITEMAD+. D. G. O. was supported by FPI grant BES-2013-066486. B. P. and A. A.-G. acknowledge the Air Force of Scientific Research for support under award: FA9550-12-1-0046. A.A.-G. acknowledges the Army Research Office under Award: W911NF-15-1-0256 and the Defense Security Science Engineering Fellowship managed by the Office of Naval Research.

\appendix

\section{\label{app:Unitary}Waves from Optical Transformations}

\begin{figure}[t]
\centering \includegraphics[width=0.8\linewidth]{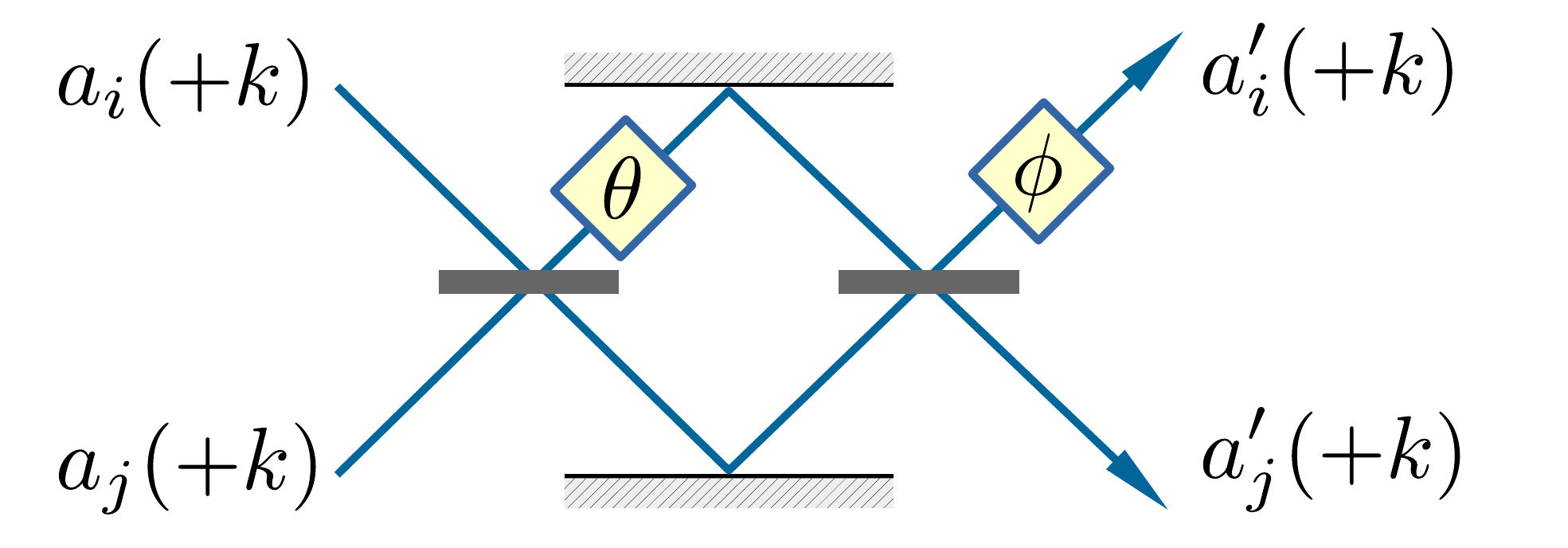}
\caption{A two-port interferometer built with two 50-50 beam splitters and two phase shifters,
$\theta$ and $\phi$, implements the most general unitary transformation from input modes $(a_i,a_j)$
into output modes $(a^\prime_m,a^\prime_n)$.}
\label{fig:Transformation}
\end{figure}

An arbitrary $M\times M$ optical transformation $U_{mn}$ may be decompose into a series of at most $M^2$ interferometers and phase shifters~\cite{Clements2016a,Reck1994a} of the form shown in Fig.~\ref{fig:Transformation}. Assuming this optical interpretation of the circuit, we will find that the $l$-th operation will couple $m$-th and $n$-th modes through a unitary transformation
\begin{equation}
\left(\begin{matrix} a^\prime_m\left(+k\right) \\ a^\prime_n\left(+k\right) \end{matrix}\right) =
U^{}_{mn}\left(+k\right)
\left(\begin{matrix} a^{}_m\left(+k\right) \\ a^{}_n\left(+k\right) \end{matrix}\right) \comma
\end{equation}
that, following the conventions in Ref.\ \cite{Reck1994a}, depends on two angular parameters
\begin{align} \label{eq:Unitary-2}
U_{l} \left( \phi_l,\theta_l \right) &:= \left( \begin{matrix}
\sin \left( \theta_{l} \nu_k \right) e^{i\phi_{l} \nu_k/2} &
\cos \left( \theta_{l} \nu_k \right) e^{i\phi_{l} \nu_k/2} \\
\cos \left( \theta_{l} \nu_k \right) &
-\sin \left( \theta_{l} \nu_k \right) \end{matrix} \right) \point
\end{align}
Note that the unitary transformation also depends on the momentum of the photon.

Taking as reference the unitary that is implemented for the photons that are resonant with the qubits,
$k=k_\Delta$, a general transformation will typically read
\begin{equation} \label{eq:Unitary-3}
U\left(+k\right)=\prod\limits_{n=1}^M U^{}_n \left( \phi_n,\theta_n,\nu_k \right) \point
\end{equation}
An important question is what happens to the photons propagating in the opposite direction. It is not difficult to convince oneself that by having a backwards-moving plane wave we will obtain the relation
\begin{equation}
\vec{a}\left(-k\right) = \prod\limits_{n=M}^1 U^T_n \left( \phi_n,\theta_n,\nu_k \right)
                                              \, \vec{a}^{\>\prime} \hspace{-0.15em} \left(-k\right) \comma
\end{equation}
where $U^T$ arises from the particular form of the optical transformation~\eqref{eq:Unitary-2}. Since the product of unitaries runs in an opposite order to that of~\eqref{eq:Unitary-3}, when we apply the inverse transformations to extract $\vec{a}^{\>\prime} \hspace{-0.15em}$ we recover the simple result
\begin{equation}
\vec{a}^{\>\prime} \hspace{-0.15em} \left(-k\right) =
\prod\limits_{n=1}^M  U^*_n \left( \phi_n,\theta_n,\nu_k \right) \, \vec{a}\left(-k\right)
=: U^* \hspace{-0.2em} \left(+k\right) \, \vec{a}\left(-k\right) \point
\end{equation}
In other words, we have found the relationship
\begin{equation} \label{eq:U-Symmetry}
U\left(-k\right) = U^*_{} \hspace{-0.2em} \left(+k\right) \comma
\end{equation}
which is a generalization of the relation between forward and backwards-moving waves, $\exp \left(\pm ikx \right)$, in one-dimensional waveguides.

\section{\label{app:Master-Eq-Derivation}Derivation of a Master Equation for the Open Circuit}

A master equation that describes the effective qubit dynamics generated by the Hamiltonian~\eqref{eq:H-Open} may be derived in the Markovian regime. This limit assumes that the travelling time of the photons through the optical circuit is much shorter than the spontaneous emission rate of the qubits, which is of the order of the spectral function $J\left(\omega\right) = \pi \sum_k g^2_k \delta(\omega_k-\omega)$ at the resonance point $\omega=\Delta$. We also assume a weak coupling limit $g_k \ll \Delta$, $\omega_k$. Under these approximations, a procedure similar to the one described in the Supplementary Material of
\cite{gonzalez-tudela13} may be followed.

We start from the Liouville-von Neumman equation after having performed the Born-Markov approximation:
\begin{align} \label{eq:Born-Markov}
\frac{\diff\rho_0}{\diff t} = & - \int_0^\infty \hspace{-0.5em} \diff\tau \, \text{tr}^{}_B
\left[ H_I \left(t\right), \left[ H_I \left(t-\tau\right), \rho_0\left(t\right) \rho_B \right]\right]
\\
= & + \int_0^\infty \hspace{-0.5em} \diff\tau \,
\text{tr}^{}_B \left\lbrace H_I \left(t-\tau\right) \rho_0\left(t\right)
                                \rho_B \, H_I \left(t\right) \right\rbrace + \text{h.c.}
\nonumber \\
\hspace{1em} & - \int_0^\infty \hspace{-0.5em} \diff\tau \,
\text{tr}^{}_B \left\lbrace H_I \left(t\right) H_I \left(t-\tau\right)
                            \rho_0\left(t\right) \rho_B \right\rbrace
+ \text{h.c.} \nonumber
\end{align}
Here $H_I\left(t\right)$ refers to the interacting part of the Hamiltonian~\eqref{eq:H-Open} in the interaction picture, and tr$_B$ refers to the partial trace over the bosonic degrees of freedom.

The next step consists in expanding these expressions and performing the rotating wave approximation, while assuming that the equilibrium state of the photonic degrees of freedom is close to the ground state. This yields for the first term in Eq.~\eqref{eq:Born-Markov}
\begin{align} \label{eq:Born-Markov-Term-1}
\int_0^\infty & \hspace{-0.5em} \diff\tau \, \text{tr}^{}_B \left\lbrace
H_I \left(t-\tau\right) \rho_0\left(t\right) \rho_B \, H_I \left(t\right) \right\rbrace =
\\ & \, \pi \sum\limits_{m,k} g^2_k \, \delta\left(\omega_k-\Delta\right)
              \hspace{-1.15em} \sum\limits_{\alpha \in \lbrace\inb,\out\rbrace} \hspace{-1.15em}
              \sigma_{\alpha,m}^- \, \rho_0 \, \sigma_{\alpha,m}^+ + i \, \Xi \nonumber
\\ + & \, \pi \hspace{-0.25em} \sum\limits_{m,n,k} \hspace{-0.25em}
          g^2_k \, \delta\left(\omega_k-\Delta\right) \,
          \mathcal{U}^{}_{nm} \left(k\right) \, \sigma_{\inb,m}^- \, \rho_0 \, \sigma_{\out,m}^+
          + \text{h.c.} \nonumber
\end{align}
where $\Xi \equiv \Xi \left[\rho_0,\ldots\right]$ is a complicated functional which is Hermitian, and therefore disappears when adding the term~\eqref{eq:Born-Markov-Term-1} to its complex conjugate in~\eqref{eq:Born-Markov}.

Following the same procedure with the second term in Eq.~\eqref{eq:Born-Markov} gives a more complicated
contribution
\begin{align} \label{eq:Born-Markov-Term-2}
\int_0^\infty & \hspace{-0.5em} \diff\tau \, \text{tr}^{}_B \left\lbrace
H_I \left(t\right) H_I \left(t-\tau\right) \rho_0\left(t\right) \rho_B \right\rbrace =
\\ & \pi \sum\limits_{m,k} g^2_k \, \delta\left(\omega_k-\Delta\right)
         \hspace{-1.15em} \sum\limits_{\alpha \in \lbrace\inb,\out\rbrace} \hspace{-1.15em}
         \sigma_{\alpha,m}^+ \, \sigma_{\alpha,m}^- \, \rho_0 \nonumber
\\ + & \, \pi \hspace{-0.25em} \sum\limits_{m,n,k} \hspace{-0.25em}
              g^2_k \, \delta\left(\omega_k-\Delta\right) \, U^{}_{nm} \left(k\right)
                    \, \sigma_{\inb,m}^+ \, \sigma_{\out,m}^- \, \rho_0
              + \text{h.c.} \nonumber
\\ - & \, i \, \text{PV} \left\lbrace \sum\limits_{m,k} \frac{g_k^2}{\omega_k-\Delta} \right\rbrace
               \hspace{-0.25em} \sum\limits_{\alpha \in \lbrace\inb,\out\rbrace} \hspace{-1.15em}
               \sigma_{\alpha,m}^+ \, \sigma_{\alpha,m}^- \, \rho_0 \nonumber
\\ - & \, i \, \text{PV} \left\lbrace \sum\limits_{m,n,k} \frac{g_k^2}{\omega_k-\Delta} \;
                                      U^{}_{nm} \left(k\right)
                                      \sigma_{\inb,m}^+ \, \sigma_{\out,m}^- \, \rho_0 + \text{h.c.}
                         \right\rbrace \nonumber
\\ - & \, i \, \text{PV} \left\lbrace \sum\limits_{m,k} \frac{g_k^2}{\omega_k+\Delta} \right\rbrace
               \hspace{-0.25em} \sum\limits_{\alpha \in \lbrace\inb,\out\rbrace} \hspace{-1.15em}
               \sigma_{\alpha,m}^- \, \sigma_{\alpha,m}^+ \, \rho_0 \nonumber
\\ - & \, i \, \text{PV} \left\lbrace \sum\limits_{m,n,k} \frac{g_k^2}{\omega_k+\Delta} \;
                                      U^{}_{nm} \left(k\right)
                                      \sigma_{\inb,m}^- \, \sigma_{\out,m}^+ \, \rho_0 + \text{h.c.}
                         \right\rbrace \nonumber
\end{align}
where PV means that the Cauchy principal value of the integrals in momenta $k$ should be computed.

The aforementioned integrations in $k$ may be substituted by integrals in the photon frequency $\omega \equiv \omega_k$ by making use of the property~\eqref{eq:U-Symmetry} of $U\left(k\right)$, and including a density of states $\mathcal{D}\left(\omega\right) = \diff_k \, \omega_k$. After this change of variable, all the integrals can be evaluated explicitly; either as a consequence of the definition of the $\delta$ distribution or by using the Kramers-Kronig relations,
\begin{align}
\real\left\lbrace f\left(\Delta\right)\right\rbrace =  \frac{1}{\pi} \int_{-\infty}^{+\infty} &
\frac{\diff\omega}{\omega-\Delta} \> \imag\left\lbrace f\left(\omega\right)\right\rbrace
\\ \nonumber \\
\imag\left\lbrace f\left(\Delta\right)\right\rbrace = -\frac{1}{\pi} \int_{-\infty}^{+\infty} &
\frac{\diff\omega}{\omega-\Delta} \> \real\left\lbrace f\left(\omega\right)\right\rbrace
\end{align}
which hold provided that $f\left(\omega\right)$ is an analytic function in the upper half of the complex plane.

Besides a renormalization of the qubit frequencies, which does not affect the dynamics in the interaction picture, the result of integrating~\eqref{eq:Born-Markov-Term-1}--\eqref{eq:Born-Markov-Term-2} is
\begin{align} \label{eq:Master-Eq-Term-1}
\int_0^\infty & \hspace{-0.5em} \diff\tau \, \text{tr}^{}_B \left\lbrace
H_I \left(t-\tau\right) \rho_0\left(t\right) \rho_B \, H_I \left(t\right) \right\rbrace + \text{h.c.} =
\\ & \Gamma \sum\limits_m \left( \sigma_{\out,m}^- \, \rho_0 \, \sigma_{\out,m}^+
                                    + \, \sigma_{\inb,m}^- \, \rho_0 \, \sigma_{\inb,m}^+ \right) \nonumber
\\ + & \, \Gamma \sum\limits_{m,n} \, \real\left[ U^{}_{nm}\right]
                                      \sigma_{\out,n}^- \, \rho_0 \, \sigma_{\inb,m}^+
   + \text{h.c.} \, \comma \nonumber
\\ \label{eq:Master-Eq-Term-2}
\int_0^\infty & \hspace{-0.5em} \diff\tau \, \text{tr}^{}_B \left\lbrace
H_I \left(t\right) H_I \left(t-\tau\right) \rho_0\left(t\right) \rho_B \right\rbrace =
\\ & \frac{\Gamma}{2} \sum\limits_m \left( \sigma_{\out,m}^+ \, \sigma_{\out,m}^-
                                         + \, \sigma_{\inb,m}^+ \, \sigma_{\inb,m}^- \right) \rho_0 \nonumber
\\ + & \, \frac{\Gamma}{2} \sum_{m,n} U^{}_{nm} \left( \sigma_{\out,n}^+ \, \sigma_{\inb,m}^- 
                                                     + \sigma_{\inb,m}^+ \, \sigma_{\out,n}^-\right) \rho_0
\, \comma \nonumber
\end{align}
where the spontaneous emission rate parameter
\begin{equation} \label{eq:SpontaneousEmissionRate}
\Gamma = J\left(\Delta\right) = 2 \, \bar{g}_\Delta^2 \mathcal{D}\left(\Delta\right)
\end{equation}
is the natural time scale for the dipolar qubit-waveguide interaction, and $U\equiv U\left(k_\Delta\right)$.

Substituting~\eqref{eq:Master-Eq-Term-1}--\eqref{eq:Master-Eq-Term-1} into~\eqref{eq:Born-Markov} gives as a result the effective master equation~\eqref{eq:Effective-Master-Eq}, with the different contributions discussed in Sec.~\ref{sec:Open-Circuit}.

\section{\label{app:Dark-State-Decays}Crowding of Asymptotic Solutions}

The stationary solutions~\eqref{eq:Dark-States} obtained in Sec.~\ref{sec:Dark-States} are delocalized excitations that, by construction, are not dissipated according to the dynamics described by the master equation~\eqref{eq:Factorized-Master-Equation}. However, there is no prescription in these dynamics preventing a state with \emph{more than one} such an excitation from dissipating. Even though this point is rigorously true, under certain conditions the decay of these `crowded' dark states may be superseded by the typical timescale $1/\Gamma$ of the effective dynamics.

The magnitude of the decay of a crowded dark state, 
\begin{equation} \label{eq:Crowded-State}
\ket{\psi_\text{dark}\left(j_1,\ldots,j_N\right)} = \prod\limits_{\alpha=1}^N W^+_{j_\alpha} \ket{0} \comma
\end{equation}
is determined, in units of $\Gamma$, by the norm of the state resulting from the application of a collective annihilator $S^-_i$ on this state:
\begin{equation} \label{eq:Crowded-Norm} \mathcal{N} = \sqrt{
\braket{ 0 | \prod\limits_{\alpha=N}^1 \hspace{-0.1em} W^-_{j_\alpha} \,
             S^+_i S^-_i \hspace{-0.1em}
             \prod\limits_{\beta=1}^N \hspace{-0.1em} W^+_{j_\beta} \, | 0 } } \comma
\end{equation}
where it is assumed that all $j_\alpha$ indices are different.

The norm~\eqref{eq:Crowded-Norm} can be calculated by commuting the $S^-_i$ operator with the string of $W^+_{j_\beta}$ operators that lay to the right, following at the same time an identical by commuting $S^+_i$ with the $W^-_{j_\alpha}$ operators to the left. The commutator of $S^-_i$ and $W^+_{j_\alpha}$ is
\begin{equation} \label{eq:Dark-State-Commutator-1}
\left[ S^-_i, W^+_{j_\beta} \right] = \frac{1}{2} \, O^{}_{j_\beta i}
                                      \left( \sigma^z_{\out,j_\beta} - \sigma^z_{\inb,i} \right) \point
\end{equation}
is required in order to conduct these operations. We also need to know that
\begin{align} \label{eq:Dark-State-Commutator-2}
\left[ \sigma^z_{\inb,j_\alpha}, W^\pm_{j_\beta} \right] =
\pm \sqrt{2} \, O^{}_{j_\beta j_\alpha} \sigma^\pm_{\inb,j_\alpha} \comma
\\ \label{eq:Dark-State-Commutator-3}
\left[ \sigma^z_{\out,j_\alpha}, W^\pm_{j_\beta} \right] =
\mp \sqrt{2} \, \delta^{}_{j_\beta j_\alpha} \sigma^\pm_{\out,j_\alpha} \point
\end{align}

After each successive conmutation~\eqref{eq:Dark-State-Commutator-1}--\eqref{eq:Dark-State-Commutator-3} performed on~\eqref{eq:Crowded-Norm}, we get either terms with the same number of Pauli matrices or terms with one less Pauli matrix. Most of these are zero upon explicit inspection, either because $(\sigma^z_{\out,j_\beta} - \sigma^z_{\inb,i}) \ket{0} = 0$, or as a consequence of having assumed that no $j_\alpha$ index is repeated, or because the application of~\eqref{eq:Dark-State-Commutator-2} leaves a vanishing product of Pauli matrices.

The only nonvanishing terms are those which depend on quadratic powers of sets of $2$, $3$, $\ldots N$ coefficients of the $i$-th column of the unitary transformation $U$, with the rows being chosen among the dark state indices $j_\alpha$ appearing in~\eqref{eq:Crowded-State}:
\begin{align} \label{eq:Crowded-Norm-Full}
\mathcal{N} = \sqrt{ \sum\limits_{\alpha<\beta} \frac{O^2_{j_\alpha i} O^2_{j_\beta i}}{2}
              \left[ 1 - \sum\limits_{\gamma\neq\alpha,\beta} \hspace{-0.4em}
                             \frac{O^2_{j_\gamma i}}{2} \left( 1 - \dots \right) \right] } \> \point
\end{align}

When averaged over the Haar measure, the expected values of the matrix elements of $O$ are $\mathcal{E} (O^{}_{ji}) \sim \mathcal{O} (\frac{1}{\sqrt{M}})$. Counting the number of different possible combinations of $n<N$ coefficients appearing in Eq.~\eqref{eq:Crowded-Norm-Full} yields 
\begin{align} \label{eq:Crowded-Norm-Expectation}
\mathcal{E} \left( \mathcal{N}^2 \right) =
& \sum\limits_{n=2}^N \frac{\left(-1\right)^n}{2^{n-1}}
\left(\begin{array}{c}N\\2\end{array}\right)
P^{N-2}_{n-2}
\left(\begin{array}{c}n\\2\end{array}\right)
P^{n-2}_{n-2} \\
= & \sum\limits_{n=2}^N
\frac{\left(-1\right)^n}{2^{n+1}} \, \frac{N!}{\left(N-n\right)!} \,
\frac{n!}{M^n}
\, \point \nonumber
\end{align}

Where the decay parameter of these quasi-stationary dark states is given by $\Gamma \times \mathcal{E} \left( \mathcal{N} \right)$. Considering the dilute limit $M\gg N > 1$, it follows that $\mathcal{E} \left( \mathcal{N} \right) \sim M^{-1} \ll 1$. The consequence of this being that the typical timescales associated to the decay of a state with multiple dark-state excitations are much larger than the characteristic timescale $\Gamma^{-1}$ of the dynamics described by Eq.~\eqref{eq:Factorized-Master-Equation}.

\section{\label{app:boson sampling}Adiabatic Connection to Boson Sampling}

Let us start by proving that, when using harmonic oscillators and non-classical states, it is possible to adiabatically prepare a boson sampling distribution. The setup that we have in mind is a collection of $M$ input and $M$ output bosonic modes (i.e. resonators) connected between themselves through an optical transformation such as the one in Fig.~\ref{fig:Setup}. Provided that these resonators satisfy the same requirements as in Sec.~\ref{sec:Open-Circuit}, we will be able to write down an effective Hamiltonian of the form
\begin{align}
\label{eq:H-Boson-Sampling}
H_{BS}= &
\sum_i \tilde{\Delta} \left( a_{\inb,n}^\dagger a^{}_{\inb,n} + a_{\out,n}^\dagger a^{}_{out,n} \right)
\nonumber \\ + &
\sum\limits_{m,n} J_{n,m} \left( a^\dagger_{\out,n} a^{}_{\inb,m} +\text{h.c} \right)
\\ + &
\epsilon \sum_n \left( \left[ 1 - \lambda\left(t\right) \right] a_{\out,n}^\dagger a^{}_{\out,n}
                     + \lambda\left(t\right) a_{\inb,n}^\dagger a^{}_{\inb,n}\right) \point \nonumber
\end{align}
Since $J_{n,m}\propto U_{n,m}$, we may define new collective operators,
\begin{equation}
c^{}_{\out,m} = \sum\limits_n U^{}_{mn} \, a^{}_{\out,n},
\end{equation}
which diagonalize the previous Hamiltonian
\begin{align}
H_{BS} = &
\sum\limits_i \left( \Delta_\inb \left(t\right) a^\dagger_{\inb,n} a^{}_{\inb,n} +
\Delta_\out\left(t\right) a_{\out,n}^\dagger a_{\out,n} \right) \\ + & \,
\delta \sum\limits_m \left( c^\dagger_{\out,n} a^{}_{\inb,m} + \text{h.c} \right) \comma \nonumber
\end{align}
where we have introduced $\Delta_\inb = \tilde{\Delta} + \epsilon \, \lambda\left(t\right)$, $\Delta_\out\left(t\right) = \tilde{\Delta} + \epsilon \left[ 1 - \lambda\left(t\right) \right]$. It is now rather simple to apply the adiabatic theorem to each of the $M$ local Hamiltonians that connect $a_{in,m}$ to the corresponding $c_{out,m}$. The result is that by switching off $\lambda\left(t\right)$ in a time $T \gg 1/\left\vert\delta\right\vert$, and provided $\left\vert\epsilon\right\vert \gg \left\vert\delta\right\vert$, we will adiabatically transfer the state $a_{\inb,m}^\dagger\ket{0}$ onto an output state $c_{\out,m}^\dagger\ket{0}$. This way, if we start with $N$ excitations
\begin{equation}
\ket{\phi\left(0\right)} = \prod\limits_{m=1}^N a^\dagger_{\inb,m} \ket{0} \comma
\end{equation}
the final state will be, up to small corrections, the output of the interferometer with $N$ input photons
\begin{equation}
\ket{\phi\left(T\right)} \simeq \prod\limits_{m=1}^N \sum\limits_n U^{}_{nm} \, a^\dagger_{\out,n} \ket{0} \point
\end{equation}

It remains to be proven that we achieve a similar state when using the spin Hamiltonian~\eqref{eq:H-Spin} and an input state
\begin{equation}
\ket{\psi(0)}=\prod_m \sigma^+_m\ket{0} \point
\end{equation}
For this we invoke the result in Ref.~\cite{Peropadre2015b} which establishes that the distance between the hard-core boson state $\ket{\psi\left(t\right)}$, evolved under \eqref{eq:H-Spin}, and the soft-boson state $\ket{\phi\left(t\right)}$ evolved under \eqref{eq:H-Boson-Sampling} is bounded by
\begin{equation}
\left\Vert \psi-\phi \right\Vert \leq
\int_0^T \hspace{-0.45em} \diff t \, \left\Vert Q \, H_{BS} \, P_\text{1pair} \right\Vert_2
                                     \left\Vert P_\text{1pair} \, \phi\left(t\right) \right\Vert_2 \comma
\end{equation}
where $Q$ and $P_{1pair}$ project onto the hard-core subspace and the space of states with at most one bunched mode. Therefore, it can be concluded that
\begin{equation}
\left\Vert \psi-\phi \right\Vert \leq \mathcal{O} \left( T\frac{N^2}{\sqrt{M}} \right) \comma
\end{equation}
so that it becomes possible to decrease the error arbitrarily by either making the system more dilute or adjusting the evolution time.

\end{document}